\documentstyle[preprint,epsfig,aps]{revtex}
\begin{document}
\draft
\preprint{\begin{tabular}{r}
{\bf hep-ph/0004020}\\
YUMS 00-03,~~~KIAS-P00007
\end{tabular}}

\title{Neutrino Masses and Leptonic CP Violation}
\author{Sin Kyu Kang\footnote{skkang@ns.kias.re.kr}}
\address{School of Physics, Korea Institute for Advanced Study, Seoul
130-012, Korea}
\author{C. S. Kim \footnote{kim@kimcs.yonsei.ac.kr,~~
http://phya.yonsei.ac.kr/\~{}cskim} and
 J. D. Kim \footnote{jdkim@theory.yonsei.ac.kr}}
\address{
Physics Department and IPAP, Yonsei University, Seoul 120-749, Korea \\ 
}


\maketitle     

\begin{abstract}
\noindent
Neutrino oscillations as solutions of the solar neutrino problems
and the atmospheric neutrino deficits may restrict neutrino
mass squared differences and mixing angles in three-neutrino 
mixing scheme. Currently we have several solutions depending on
the interpretations of the solar neutrino problems.  Combining 
the neutrino oscillation solutions and a mass matrix ansatz, 
we investigated the neutrino mass bounds and found possible 
leptonic CP-violating rephasing-invariant quantity 
$J^{l}_{\mbox{\tiny CP}} \leq 0.012$ for large mixing angle MSW and 
just-so vacuum oscillations solutions, and
$J^{l}_{\mbox{\tiny CP}} \leq 0.0013$ for small mixing angle MSW solution.
\end{abstract}


\section{Introduction}
\label{sec:intro}
The quark masses and the related flavor mixings are the most 
intriguing riddles to be understood in the Standard
Model (SM).  Within the SM, the quark masses and flavor mixing 
angles are not predictable.  The flavor mixing angles arise 
since the quark states for which the weak interaction is diagonal 
are not mass eigenstates.  Moreover, degenerate quarks of a given 
charge render the flavor mixing angles physically meaningless. 
Thanks to that fact, we get a hint that the flavor mixing angles 
can be related to the elements of quark mass matrix. As an attempt 
to provide any relationship between the flavor mixing angles and 
the elements of quark mass matrix,  mass matrix ansatz has been 
suggested~\cite{fritz,kang}.  With the help of a mass matrix ansatz, 
we may predict some free parameters in the SM.

On the other hand, recent neutrino experimental results and cosmological
observations provide evidence for non-zero neutrino masses and 
the possible lepton  flavor mixings. Then, the SM has to be enlarged 
and we have more free parameters to describe all fermion masses and 
their mixing angles.  In this case, one may also reduce the number of 
free parameters by using separate lepton mass matrix ansatz. 
If it is possible to provide some quark-lepton symmetry in the quark 
and lepton mass matrices, one may reduce the number of free parameters 
much more.  In recent work~\cite{our}, we showed that the hierarchical quark 
mixing pattern as well as bimaximal lepton mixing pattern can arise 
from one single particular mass matrix based on the permutation symmetry with 
suitable breaking.  Remarkably those different mixing patterns 
could be generated by using the same texture of the mass matrices 
for quarks and leptons but with different hierarchies.
However, although non-zero neutrino masses and mixings can be interpreted
as a solution to the solar~\cite{bahcall} and the atmospheric~\cite{SK} 
neutrino anomalies, the present neutrino experimental results do not pin down 
the values of neutrino masses and mixing angles in three-neutrino
oscillation scheme. Moreover, the solution for the solar neutrino deficit 
may be either small or large mixing with different mass squared differences
depending on whether we consider the matter effect (Mikheyev, Smirnov,
and Wolfensten (MSW) effect)~\cite{msw,msw2} or not. 
Thus, one can at best estimate the hierarchy of neutrino mass patterns and
their mixings  case by case. In this work, we show in more detail 
how the lepton flavor mixing matrix can be obtained from a specific form 
of lepton mass matrix by assuming quark-lepton symmetry for the fermion 

small mixing angle solutions for the solar neutrino deficit combined 
with the large mixing solution for the  Super-Kamiokande
atmospheric neutrino anomaly.

In addition, it will be very interesting to study the possible
CP violation in the lepton sector, which arises due to the non-vanishing
CP phase in the flavor mixing matrix~\cite{cpr}. To do this, we will calculate 
the invariant leptonic CP violating quantity $J^{l}_{\mbox{\tiny CP}}$ 
\cite{jar} from the phenomenological lepton flavor mixing matrix. 
As will be shown later, the invariant quantity,
$J^{l}_{\mbox {\tiny CP}}$, depends on the neutrino masses as well as the CP
phase. {}From the estimate of the neutrino mass bounds based on the neutrino
experimental results, we will provide
the possible range of $J^l_{\mbox{\tiny CP}}$.
\section{Neutrino Mixing Matrix with a CP Violating Phase}
\label{sec:mix}
Let us begin by assuming that the form of lepton mass matrix can be
derived from the mass matrix ansatz based on the permutation symmetry 
with suitable breaking which is used in the quark sector~\cite{our}. 
As shown in Ref.~\cite{kang}, the mass matrix has the following form:
\begin{equation}
M_H = \left(
   \begin{array}{ccc}
     0    &   A   & 0 \\
     A    &   D   & B \\
     0    &   B   & C
   \end{array}
   \right).
\end{equation}
The parameters $A,B,C$ and $D$ can be expressed in terms of the fermion
mass eigenvalues and one free parameter $\epsilon^l$.
One can take the mass eigenvalues to be $-m_1, m_2$ and $m_3$
with the following three conditions:
\begin{eqnarray}
Tr(M_H) &=& -m_1+m_2+m_3,  \nonumber   \\
Det(M_H) &=& -m_1m_2m_3, \nonumber\\
{\rm and}~~~~Tr(M_H^2) &=& m_1^2+m_2^2+m_3^2.
\end{eqnarray}
The sign of the fermion mass is irrelevant since it can be changed
by a chiral transformation. {}From those relations, we obtain 
the following form of fermion mass matrix:
\begin{equation}
M= \left(
   \begin{array}{ccc}
     0    &     \sqrt{\frac{m_1 m_2 m_3}{m_3-\epsilon^l}} & 0 \\
     \sqrt{\frac{m_1 m_2 m_3}{m_3-\epsilon^l}} & m_2-m_1+\epsilon^l
                           & \zeta(m_2-m_1+\epsilon^l) \\
     0 & \zeta(m_2-m_1+\epsilon^l)  & m_3-\epsilon^l
   \end{array}
   \right) ,
\label{mm}
\end{equation}
in which  the analytic relation between two parameters
$\epsilon^l$ and $\zeta$ is given \cite{our} by
\begin{equation}
\zeta^2=
   \frac{\epsilon^l (m_3-m_2+m_1-\epsilon^l)(m_3-\epsilon^l)-
        \epsilon^l m_1m_2}
        {(m_3-\epsilon^l)(m_2-m_1+\epsilon^l)^2} .
\label{omega2}
\end{equation}
With the help of the analytic form of the orthogonal matrix $U$ presented 
in Ref.~\cite{our}, the real symmetric mass matrix $M$ can be diagonalized. 
The mass matrices for charged leptons and neutrinos have the same form 
of the mass matrix (3). In particular, notice that the parameter $\epsilon^l$
will be taken to be identical in both the charged lepton mass matrix and 
the neutrino mass matrix, and will be determined from the neutrino 
experimental results. However, the parameters $\zeta$ are different 
in the two mass matrices because they depend on their fermion masses.
Then, the neutrino mass matrix $M_{\nu}$ and charged lepton mass matrix 
$M_l$ can be brought to diagonal forms by the real unitary matrices 
$U_{\nu}$ and $U_l$,
\begin{eqnarray}
U_{\nu} M_{\nu} U_{\nu}^{\dagger} &=& \mbox{diag}(-m_1,m_2,m_3),\nonumber \\
U_l M_l U_l^{\dagger} &=& \mbox{diag}(-m_e,m_{\mu},m_{\tau}), \nonumber
\end{eqnarray}
where $m_1,m_2,$ and $m_3$ are neutrino masses from now on.
The lepton flavor mixing matrix $V^l_{\mbox{\tiny CKM}}$ 
is related to $U_{\nu}$ and $U_l$ as follows:
\begin{equation}
V^l_{\mbox{\tiny CKM}}=PU_lP^{-1} U_{\nu}^T ,
\label{vlepton}
\end{equation}
where the phase matrix is $P={\rm diag}(e^{i\delta^l},1,1)$.
More generally, we can also take the phase matrix, $P$, as
${\rm diag}(e^{i\delta_1},e^{i\delta_2},e^{i\delta_3})$.
One may eliminate the phase $\delta_3$ by a phase transformation of fields.
Because of the hierarchy of the charged lepton masses, 
Eq.~(\ref{vlepton}) contains only the combination of phases in the form,
$\delta_1-\delta_2$, which will be identified as $\delta^l$.
To see easily how the lepton mixing pattern is related to
the lepton mass hierarchy, first of all, we present the lepton flavor 
mixing matrix {\it in the leading approximation}.
In the next section, the exact form derived from Eq. (5) will be used to 
determine the magnitudes of the elements of the mixing matrix.

Since the charged lepton family has pronounced mass hierarchy
$m_e\ll m_{\mu} \ll m_{\tau}$, the charged lepton mass matrix can be presented
in the approximate form as
\begin{equation}
M_l \simeq
\left(
\begin{array}{ccc}
0   &   \sqrt{m_e m_{\mu}}   & 0 \\
\sqrt{m_e m_{\mu}}  &  m_{\mu}  &  \sqrt{\epsilon^l m_{\tau}}  \\
0   & \sqrt{\epsilon^l m_{\tau}}   &   m_{\tau}
\end{array} \right).  
\end{equation}
{}From the unitary transformation, we obtain the approximate form of the matrix
$U_l$ as follows
\begin{equation}
U_l \simeq
\left(
\begin{array}{ccc}
1   &   -\sqrt{\frac{m_e}{m_{\mu}}}   & 0 \\
\sqrt{\frac{m_e}{m_{\mu}}}  &  1  &  -\sqrt{\frac{\epsilon^l}{m_{\tau}}}  \\
0   & \sqrt{\frac{\epsilon^l}{m_{\tau}}}   &   1
\end{array}
\right) ,
\end{equation}
where we assumed that the parameter 
$\epsilon^l \ll m_{\tau}$.
On the other hand,  the neutrino mass matrix can be obtained from
the mixing pattern among three neutrinos and their mass hierarchy.
As shown in Ref. ~\cite{our},
the large mixing between $\nu_{\mu}$ and $\nu_{\tau}$,
which is a solution for the atmospheric neutrino anomaly, 
can be achieved by taking $\epsilon^l\simeq m_3/2$ and $m_1, m_2 \ll m_3$
in Eq.~(\ref{mm}).
We also note that the large (small) mixing between $\nu_e$ and $\nu_{\mu}$,
which is a solution for the solar neutrino deficit,
can be obtained by taking $m_1\simeq m_2 ~~(m_1<<m_2)$.
Keeping the next-to-leading order, the neutrino mass matrix, Eq.~(\ref{mm}), 
becomes
\begin{equation}
M_{\nu} \simeq
\left(
\begin{array}{ccc}
     0    &   \sqrt{2m_1 m_2}   & 0 \\
     \sqrt{2m_1 m_2}    &   \frac{m_3}{2}\left(1+2\frac{m_2-m_1}{m_3}\right)
      &   \frac{m_3}{2}\left(1-\frac{m_2-m_1}{m_3}\right)  \\
      0   &   \frac{m_3}{2}\left(1-\frac{m_2-m_1}{m_3}\right)  
          & \frac{m_3}{2}
   \end{array}
   \right) ,
\end{equation}
and the form of  $U_{\nu}$ is approximately given by
\begin{equation}
U_{\nu} \simeq
\left(
\begin{array}{ccc}
w_2 \left(1+\frac{m_1}{2m_3}\right) &
-\frac{w_1}{\sqrt{2}}
      \left(1+\frac{m_1}{2m_3}\right) &
\frac{w_1}{\sqrt{2}}
      \left(1-\frac{m_2}{m_3}-\frac{m_1}{2m_3}\right) \\
w_1\left(1-\frac{m_2}{2m_3}\right) &
\frac{w_2}{\sqrt{2}} \left(1-\frac{m_2}{2m_3}\right) &
-\frac{w_2}{\sqrt{2}}
      \left(1+\frac{m_2}{2m_3}+\frac{m_1}{m_3}\right) \\
\sqrt{\frac{m_1m_2}{m_3^2}} &
\frac{1}{\sqrt{2}}
      \left(1+\frac{m_2-m_1}{2m_3}\right) &
\frac{1}{\sqrt{2}}
      \left(1-\frac{m_2-m_1}{2m_3}\right) \\
\end{array}
\right),
\label{nmix}
\end{equation}
where  
$$ 
w_1 \equiv \sqrt{\frac{m_1}{m_1+m_2}}~~~
{\rm and}~~~
w_2 \equiv \sqrt{\frac{m_2}{m_1+m_2}}
~~~{\rm with}~~~ w_1^2+w_2^2=1 .
$$ 
{}From Eqs. (5), (7) and (9), the lepton flavor mixing matrix is expressed 
in the leading order  in terms of the lepton masses, $w_1$, $w_2$,
and the CP phase $\delta^l$:
\begin{equation}
V_{\mbox {\tiny CKM}}^{l} \simeq
\left(
\begin{array}{ccc}
w_2+w_1\sqrt{\frac{m_e}{2m_{\mu}}}e^{i\delta^l}   &
w_1-w_2\sqrt{\frac{m_e}{2m_{\mu}}}e^{i\delta^l}   &
-\sqrt{\frac{m_e}{2m_{\mu}}}e^{i\delta^l} \\
-\frac{w_1}{\sqrt{2}}+w_2\sqrt{\frac{m_e}{m_{\mu}}}e^{-i\delta^l} &
\frac{w_2}{\sqrt{2}}+w_1\sqrt{\frac{m_e}{m_{\mu}}}e^{-i\delta^l} &
\frac{1}{\sqrt{2}} \\
\frac{w_1}{\sqrt{2}} & -\frac{w_2}{\sqrt{2}} & \frac{1}{\sqrt{2}} 
\end{array}
\right) .
\label{mix1}
\end{equation}
The CP-violating rephasing-invariant quantity,
$J^l_{\mbox{\tiny CP}}$, is presented by
\begin{eqnarray}
J^l_{\mbox{\tiny CP}}
\equiv Im[V^l_{11}V^{l\ast}_{12}V^{l\ast}_{21}V^l_{22}]
= w_1w_2\frac{1}{2}\sqrt{\frac{m_e}{2m_{\mu}}}\sin\delta^l .
\label{jcp1}
\end{eqnarray}

Now let us express the lepton flavor mixing matrix in the standard 
parametrization \cite{stand}. As is well known, in the quark sector
the standard parametrization is given by
\begin{eqnarray}
V^{l}_{\mbox {\tiny CKM}}=\left(
\begin{array}{ccc}
 c_{12}c_{13}       &  s_{12}c_{13}     &  s_{13}e^{-i\delta_{13}}  \\
-s_{12}c_{23}-c_{12}s_{23}s_{13}e^{i\delta_{13}} &
        c_{12}c_{23}-s_{12}s_{23}s_{13}e^{i\delta_{13}} & s_{23}c_{13}  \\
 s_{12}s_{23}-c_{12}c_{23}s_{13}e^{i\delta_{13}} &
       -c_{12}s_{23}-s_{12}c_{23}s_{13}e^{i\delta_{13}} & c_{23}c_{13} \\
\end{array}
\label{stan}
\right),
\end{eqnarray}
where $s_{ij},c_{ij}$ stand for $\sin{\theta_{ij}}$ and 
$\cos{\theta_{ij}}$, respectively.  
One can then relate the elements 
of the mixing matrix in the standard parametrization to the elements of 
the flavor mixing matrix~(\ref{mix1}) by using the fact that 
the magnitudes of the mixing matrix elements and the Jarlskog 
rephasing-invariant quantity, $J^l_{\mbox{\tiny CP}}$,
are independent of the parametrization.  
Then, the mixing angles, $\theta_{ij}$, can be expressed by
\begin{eqnarray}
 \tan \theta_{12} &=& \frac{|V^l_{12}|}{|V^l_{11}|} 
\simeq \sqrt{\frac{w_1^2 + w_2^2 (m_e/2m_{\mu}) -2 w_1 w_2
       \sqrt{m_e/2m_{\mu}}\cos \delta^l}
{w_2^2 + w_1^2 (m_e/2m_{\mu}) +2 w_1 w_2 \sqrt{m_e/2m_{\mu}}\cos \delta^l}} ,
\label{tan12}  \\
\sin \theta_{13} &=& |V^l_{13}| ~~\simeq~\sqrt{\frac{m_e}{2m_{\mu}}} ,
\label{sin13}  \\
\tan \theta_{23} &=& \frac{|V^l_{23}|}{|V^l_{33}|} ~~\simeq~1 .
\label{tan23}
\end{eqnarray}
The magnitude of $V^{l}_{13}$ can be constrained by the CHOOZ 
experimental results \cite{chooz} and it turns out to be small, 
i.e., $|V^{l}_{13}|\leq 0.22$.  Then, the lepton mixing matrix~(\ref{stan}) 
is approximately written as
\begin{equation}
V^l_{\mbox {\tiny CKM}} \simeq
\left(
\begin{array}{ccc}
 c_{12}      &  s_{12}     &  s_{13}e^{-i\delta_{13}}  \\
 -s_{12}c_{23} & c_{12}c_{23} & s_{23}  \\
  s_{12}s_{23} & -c_{12}s_{23} & c_{23}
\end{array}
\right), 
\label{mixst}
\end{equation}
where $\theta_{12}$ can be either large or small,
depending on the solar neutrino oscillation solution.

Taking $ w_1 \simeq w_2 \simeq \sqrt{1/2}$ (i.e., $m_1 \simeq m_2$),
one can obtain the {\it nearly} bimaximal mixing, 
which corresponds to
$$ \theta_{12}\simeq\theta_{23}\simeq\pi/4~
~~({\it i.e.}~c_{12}=s_{12}=c_{23}=s_{23}=1/\sqrt{2}),$$
$$\sin\theta_{13}\simeq\sqrt{m_e/2m_\mu}~~~{\rm and}~~~
\delta_{13}\simeq\delta^l.$$
Then, the mixing matrix can be written as follows:
\begin{eqnarray}
V^l_{\mbox {\tiny CKM}}  \simeq
\left(
\begin{array}{ccc}
\frac{1}{\sqrt{2}} & \frac{1}{\sqrt{2}} & 
     \frac{1}{\sqrt{2}}\sqrt{\frac{m_e}{m_\mu}}e^{-i\delta^l} \\
-\frac{1}{2} & \frac{1}{2} & \frac{1}{\sqrt{2}} \\
\frac{1}{2} & -\frac{1}{2} & \frac{1}{\sqrt{2}} \\
\end{array}
\right) .
\label{bimx}
\end{eqnarray}
Note that our mixing matrix contains possible CP-violating phase
with nonzero but small
$$
|V^l_{13}| ~(=\sin\theta_{13}) \simeq
\frac{1}{\sqrt{2}}\sqrt{\frac{m_e}{m_{\mu}}} \simeq 0.05, 
$$
which is still consistent
with the bound obtained from present CHOOZ experiment \cite{chooz}.
However, if it  turns out that 
the neutrino mixing pattern is {\it exact} bimaximal mixing as suggested 
in Refs.~\cite{peccei,alta2}, the element $|V^l_{13}|$ would become 
exactly zero and then we could not see any CP violation effects 
in the leptonic sector.

In the limit of small mass ratio $m_1/m_2$, which
corresponds to the small mixing angle solution of the solar neutrino 
oscillation, the lepton mixing matrix~(\ref{mixst}) becomes
\begin{equation}
V_{\mbox {\tiny CKM}}^{l} \simeq
\left(
\begin{array}{ccc}
 c_{12}      &  s_{12}     &  \frac{1}{\sqrt{2}} 
     \sqrt{\frac{m_e}{m_\mu}}e^{-i\delta^l} \\
 -\frac{s_{12}}{\sqrt{2}} & \frac{c_{12}}{\sqrt{2}} & 
 \frac{1}{\sqrt{2}}  \\ 
 \frac{s_{12}}{\sqrt{2}} & -\frac{c_{12}}{\sqrt{2}} & 
 \frac{1}{\sqrt{2}}  \\ 
\end{array}
\right) ,
\label{mix3}
\end{equation}
where $\tan \theta_{12}\simeq \sqrt{\frac{m_1}{m_2}+
\frac{m_e}{2m_\mu}-\sqrt{\frac{m_1m_e}{2m_2m_\mu}}\cos\delta^l}  $.
We note that the mixing angle $\theta_{12}$ is correlated with the
phase $\delta^l$ in this case.

\section{Neutrino Masses and Leptonic CP Violation}
In order to determine the magnitudes of the elements of the neutrino mass
matrix and the possible range of  CP violation, we have to obtain 
numerical values of the neutrino mass eigenvalues.
Although we cannot obtain those values exactly, some possible ranges
of the neutrino mass eigenvalues can be estimated from the recent
experimental results by taking reasonable assumptions.

Most data on neutrino mixings are presented in the two-neutrino
scheme. The results are expressed in $(\Delta m^2, \sin^2{2\theta})$ plot.
With a proper approximation we can use the data on solar and 
atmospheric neutrino oscillations to make analyses for the three-neutrino 
scheme~\cite{bilenky}.  First we assume that the solar neutrino problems 
are solved by two-neutrino vacuum oscillations of 
$\nu_e\leftrightarrow\nu_{\mu}$. The survival probability for 
solar electron-neutrino in two-neutrino mixing scheme is given by
\begin{equation}
P(\nu_e\rightarrow\nu_e)=1-\sin^2{2\theta_{sol}}
     \sin^2\left(\frac{\Delta m^2_{sol}}{4}
                 \frac{L}{E}\right)  .
\label{pesur2}
\end{equation} 
In the case of $m^2_3 L/E \gg 1$ and $|V^l_{13}| \ll 1$, 
the survival probability for electron-neutrino
in the three-neutrino mixing scheme may be written as
\begin{equation}
P(\nu_e\rightarrow\nu_e)\simeq 1-4|V^l_{11}V^l_{12}|^2
     \sin^2\left(\frac{\Delta m^2_{21}}{4}
                 \frac{L}{E}\right)  .
\label{pesur3}
\end{equation}
Therefore, the mass squared difference and mixing angle 
for solar neutrino analysis in the two-neutrino scheme are
related to the mass squared difference and the standard
mixing angle, $\theta_{12}$, in the three-neutrino scheme:
\begin{equation}
\Delta m^2_{sol} \simeq \Delta m^2_{21}, \;\;\;
\theta_{sol} \simeq \theta_{12} .
\label{parasol}
\end{equation}
If we consider the matter effect in the Sun,
the survival probability for electron neutrinos, 
Eqs. (\ref{pesur2},\ref{pesur3}), is  no longer valid.
However, we can still make the connections between two- and three-neutrino
oscillation parameters by Eq.~(\ref{parasol}) in this situation; 
the mixing angle $\theta_{13}$ is small, i.e. $|V^l_{13}| \ll 1$, and the 
third neutrino mass $m_3$ is so large that just one resonance conversion 
between $m_1$ and $m_2$ neutrino mass states can take place. In this case 
the three-neutrino mixing scheme may effectively be reduced to the two-neutrino 
mixing scheme and Eq.~(\ref{parasol}) remains valid. 
If three neutrino masses are degenerate such that
the second resonance conversion could not be negligible, or
the mixing element $|V^l_{13}|$ is large, then we have to analyze 
neutrino mixing data within full three-neutrino scheme.

Likewise, we can consider the atmospheric neutrino case.
The atmospheric neutrino deficit seems to be explained by
oscillation between $\nu_{\mu}$ and $\nu_{\tau}$ with large mixing.
The survival probability for atmospheric muon-neutrino
in the two-neutrino mixing scheme is given by
\begin{equation}
P(\nu_{\mu}\rightarrow\nu_{\mu})=1-\sin^2{2\theta_{atm}}
     \sin^2\left(\frac{\Delta m^2_{atm}}{4}
                 \frac{L}{E}\right)  .
\end{equation}
In the case of $(m^2_2-m^2_1) L/E \ll 1$,
we can write
the survival probability for muon-neutrino 
in the three-neutrino mixing scheme as follows:
\begin{equation}
P(\nu_{\mu}\rightarrow\nu_{\mu})\simeq 
      1-4|V^l_{23}|^2(1-|V^l_{23}|^2)
     \sin^2\left(\frac{\Delta m^2_{31}}{4}
                 \frac{L}{E}\right)  .
\end{equation}
The mass squared difference and mixing angle for atmospheric neutrino    
analysis in the two-neutrino scheme are related to the mass squared
difference and the standard mixing angle, $\theta_{23}$, 
in the three-neutrino scheme:
\begin{equation}
\Delta m^2_{atm} \simeq \Delta m^2_{31} , \;\;\;
\theta_{atm} \simeq \theta_{23} .
\label{paraatm}
\end{equation}

Recent Super-Kamiokande experiments~\cite{SK} show
evidence for oscillation of atmospheric neutrinos. 
The data exhibit a zenith angle dependent deficit of
muon neutrinos, which is consistent with predictions
based on the two-flavor 
$\nu_{\mu}\leftrightarrow\nu_{\tau}$ oscillations.
At 90\% confidence level the mass squared difference and mixing
angle are
\begin{eqnarray}
5\times 10^{-4} &< \Delta m^2_{atm} <& 6\times 10^{-3} {\rm eV}^2 ,\\
0.82 &< \sin^2{2\theta_{atm}}\leq & 1  .  
\label{atmang} 
\end{eqnarray}
The best fit values are $\Delta m^2_{atm}\simeq 2.2\times 10^{-3}$
eV$^2$ and $\sin^2{2\theta_{atm}}=1$.  {}From the mixing
matrix~(\ref{vlepton})  the mixing angle $\theta_{23}$ is expressed in terms
of the   neutrino masses and the parameter $\epsilon^l$ as 
\begin{equation}
\tan{\theta_{23}} \simeq \sqrt{\frac{\epsilon^l}{m_3-\epsilon^l}} ~.
\end{equation}
We can constrain the ratio, $\epsilon^l/m_3$,
from Eqs.~(\ref{paraatm}) and~(\ref{atmang}) 
\begin{equation}
0.28 \leq \frac{\epsilon^l}{m_3} \leq 0.5 ~.
\label{rat1}
\end{equation}
If we take the best fit value for the mass squared difference and
assume the mass hierarchy of $m_1,m_2 \ll m_3$,  we conclude that
\begin{eqnarray}
m_3 \simeq \sqrt{\Delta m^2_{atm}} &\simeq &
                4.7\times 10^{-2} {\rm eV} ,\\
1.3\times 10^{-2} &\leq \epsilon^l \leq& 2.4\times 10^{-2} {\rm eV} .
\end{eqnarray}

As is well known, there are three oscillation solutions of 
the solar neutrino problems in the two-neutrino scheme~\cite{bahcall}:
\begin{itemize}
\item{the Large Mixing Angle (LMA) MSW}:
\begin{eqnarray}
5\times 10^{-6} &\leq \Delta m^2_{sol} \leq &
               4\times 10^{-5} {\rm eV}^2 , 
\label{solmaslma}               \\
0.4 &\leq \sin^2{2\theta_{sol}}\leq & 0.9 .
\label{solanglma}
\end{eqnarray}
\item{Vacuum Oscillations (VO)}:
\begin{eqnarray}
5\times 10^{-11} &\leq \Delta m^2_{sol} \leq & 
            10^{-10} {\rm eV}^2,
\label{solmasvo}   \\
0.67 &\leq \sin^2{2\theta_{sol}}\leq & 1 .
\label{solangvo}
\end{eqnarray}
\item{the Small Mixing Angle (SMA) MSW}:
\begin{eqnarray}
3.8\times 10^{-6} &\leq \Delta m^2_{sol} \leq & 
       10^{-5} {\rm eV}^2 ,
\label{solmassma}         \\
 3.5\times 10^{-3} &\leq \sin^2{2\theta_{sol}}\leq & 1.4\times 10^{-2} .
\label{solangsma}
\end{eqnarray}
\end{itemize}
The intervals are 95\% Confidence Level.
We will not consider the MSW solution with low mass
at $\Delta m^2_{sol}=7.9\times 10^{-8}$ eV$^2$, 
$\sin^2{2\theta_{sol}}=0.96$.
All the solutions are consistent with the predictions of
the standard solar model~\cite{bahcall} and with the observed
average event rates in 
the Chlorine (Homestake) experiments~\cite{chl}, 
Kamiokande~\cite{kamio}, 
Super-Kamiokande~\cite{Skamio}, 
Gallium (GALLEX~\cite{gal} and SAGE~\cite{sage}) experiments.
With these results from the solar neutrino oscillation solutions, we can 
investigate the bounds on the neutrino masses, $m_1,m_2$, and 
CP violation quantity,
$J^l_{\mbox{\tiny CP}}$, based on the mass matrix ansatz~(\ref{mm}).
Although it is difficult to severely constrain
the CP-violating phase from the results of solar and atmospheric experiments, 
we can get the possible ranges
of magnitude of $J^l_{\mbox{\tiny CP}}$ for a given non-zero CP phase.
To show this in detail, we will treat three cases of the solar neutrino
oscillation solutions separately.
\begin{itemize}
\item{LMA solution} \\
Recent experimental results from Super-Kamiokande seem to provide
some encouragement for considering the LMA solution of
the MSW effect~\cite{bahcall2}.
The best fit values are at $\Delta m^2_{sol}\simeq 10^{-5}{\rm eV}^2$
and $\sin^2{2\theta_{sol}}\simeq 0.6$.  The lepton flavor mixing matrix 
for this solution in the leading approximation 
is given by Eq.~(\ref{bimx}). {}From Eq.~(\ref{tan12}), 
the neutrino mixing angle $\theta_{12}$ is expressed 
in terms of the lepton masses, $m_e, m_\mu, m_1,~m_2$, and 
the CP phase $\delta^l$.
Since the LMA solution is the case of $m_1\simeq m_2$,
one may ignore relatively small terms which 
contain the ratio $m_e/m_\mu$ in Eq. (\ref{tan12}). Then, we approximately get
\begin{equation}
\theta_{sol} \simeq \theta_{12} \simeq \arctan \left({\frac{m_1}{m_2}}\right),
\label{rel2}
\end{equation}
which is in turn bounded by Eq.~(\ref{solanglma}).
Using the bounds of the mass squared difference~(\ref{solmaslma}) and 
mixing angle $\theta_{12}$, we obtain 
numerically allowed neutrino mass bounds:
\begin{eqnarray}
3.0\times 10^{-4} &\leq& m_1 \leq 2.0\times 10^{-3} {\rm eV} , \\
2.7\times 10^{-3} &\leq& m_2 \leq 1.5\times 10^{-2} {\rm eV} .
\end{eqnarray}

With the 90\% confidence limit data on neutrino oscillation parameters
we can estimate the possible ranges of the magnitude of the complex
mixing matrix elements from the numerical analysis based on the exact
form of the mixing matrix: 
\begin{equation}
|V_{\mbox {\tiny LMA}}|=
\left(
\begin{array}{ccc}
0.82 \sim 0.94 & 0.35 \sim 0.55 & 0.01 \sim 0.10 \\
0.25 \sim 0.50 & 0.55 \sim 0.77 & 0.56 \sim 0.70 \\
0.14 \sim 0.34 & 0.50 \sim 0.65 & 0.70 \sim 0.82 \\
\end{array}
\right) .
\end{equation}
{}From these results, we can calculate the quantity $J^l_{\mbox{\tiny CP}}$
as a function of the CP phase $\delta^l$.
In Fig. 1.(A), we present our prediction for the allowed range of 
$J^l_{\mbox{\tiny CP}}$ as a function of $\delta^l$, which is consistent
with the solar and atmospheric neutrino experimental results.
\item{VO solution} \\
This solution shows that there are well-separated two mass
squared difference scales, 
$\Delta m^2_{atm}\simeq 2.2\times 10^{-3}{\rm eV}^2$ and
$\Delta m^2_{sol}\simeq 8\times 10^{-11}{\rm eV}^2$.
The mixing angle $\theta_{12}$ is also determined
by  the neutrino masses $m_1$ and $m_2$ as in the case of 
LMA solution~(\ref{rel2}). {}From the constraints Eqs.~(\ref{solmasvo}) 
and~(\ref{solangvo}), we get, at best, the lower bounds 
on neutrino masses as follows:
\begin{equation}
m_1\geq 0.24\times 10^{-5}{\rm eV},  \;\;\;
m_2\geq 0.93\times 10^{-5}{\rm eV}.
\end{equation}
The limits on magnitudes of the elements of the complex
mixing matrix for this solution are
\begin{equation}
|V_{\mbox {\tiny VO}}|=
\left(
\begin{array}{ccc}
0.70 \sim 0.90 & 0.45 \sim 0.71 & 0.02 \sim 0.06 \\
0.35 \sim 0.60 & 0.47 \sim 0.75 & 0.54 \sim 0.72 \\
0.20 \sim 0.50 & 0.40 \sim 0.65 & 0.70 \sim 0.84 \\
\end{array}
\right) .
\end{equation}
Since the rephasing-invariant $J^l_{\mbox{\tiny CP}}$ given in 
Eqs.~(\ref{jcp1})
actually depends on the ratio $m_1/m_2$ and the parameter $\epsilon^l$,
we can calculate the numerical values of $J^l_{\mbox{\tiny CP}}$ with the
help of Eqs. (\ref{rat1}), (\ref{solangvo}), and (\ref{rel2}).
The results are shown in Fig. 1.(B).
As one can see from Fig. 1.(B), the allowed range of
$J^l_{\mbox{\tiny CP}}$ for the VO solution is
almost the same as that for the LMA solution.
The reason is that both solutions have almost the same neutrino mixing
angle $\theta_{12}$ and our ansatz leads to those solutions when we take
the mass hierarchy $m_1\simeq m_2$. Therefore the mixing matrix may also be
accommodated by Eq.~(\ref{bimx}) in the leading approximation
as in the case of LMA solution.
\item{SMA solution} \\
The small mixing angle $\theta_{12}$ implies small mass ratio, 
$m_1/m_2$, in Eq.~(\ref{vlepton}). Different from the above two cases, 
the lepton mixing matrix for the SMA solution in the leading approximation
is given by Eq. (\ref{mix3}).
In this case the angle depends sensitively on the phase $\delta^l$ 
as well as on the ratio $m_1/m_2$. Note that 
the expression~(\ref{rel2}) does not hold in the extreme mass
hierarchical case of $m_2 \gg m_1$, because we cannot neglect 
the $m_e/m_\mu$-terms in Eq. (\ref{tan12}). Also, the dependence of CP phase
should be taken into account when we calculate the mixing angle
$\theta_{12}\simeq \theta_{sol}$. {}From the numerical analysis based
on the exact form of the mixing matrix, 
it has been shown that the mass $m_1$ may have small value without
a lower bound in the allowed parameter space given by Eqs. 
(\ref{solmassma})-(\ref{solangsma}). 
The upper bound of mass $m_1$ depends on the value of 
CP-violating phase $\delta^l$. When $\delta^l$ is in the range of
$0<\delta^l<\pi/2$, the upper bound of $m_1$ may be up to 
$4.2\times 10^{-5}$ eV. As $\delta^l$ approaches $\pi$,
the upper bound of $m_1$ becomes smaller, around $3.0\times 10^{-7}$ eV.
{}From Eq.~(\ref{solmassma}), the lower mass bound 
of $m_2$ is to be around $1.9\times 10^{-3}$ eV.
The limits on magnitudes of the elements of the complex
mixing matrix for this solution are
\begin{equation}
|V_{\mbox {\tiny SMA}}|=
\left(
\begin{array}{ccc}
0.98 \sim 0.99 & 0.03 \sim 0.06 & 0.03 \sim 0.05 \\
0.03 \sim 0.08 & 0.70 \sim 0.84 & 0.54 \sim 0.71 \\
0.01 \sim 0.05 & 0.54 \sim 0.71 & 0.70 \sim 0.83 \\
\end{array}
\right) .
\end{equation}
Fig. 2 shows the rephasing-invariant quantity $J^l_{\mbox{\tiny CP}}$ 
for the SMA solution as function of the CP phase $\delta^l$.
The allowed range of $J^l_{\mbox{\tiny CP}}$ is presented by the hatched
region which comes from the constraints Eqs.~(\ref{solmassma}) 
and~(\ref{solangsma}). The shape of allowed region is different from 
that of large mixing of solar neutrinos. The maximum value of 
$J^l_{\mbox{\tiny CP}}$ for the SMA solution is $1.3 \times 10^{-3}$, 
while that for LMA and VO is of order $\sim 0.01$. In particular, 
as can be seen from Fig. 2, the magnitude of $J^l_{\mbox{\tiny CP}}$ is 
suppressed and severely constrained for the range of $\pi/2<\delta^l<\pi$.
Also, somewhat broad range of $J^l_{\mbox{\tiny CP}}$
for $\delta^l\sim 1$ is obtained.
\end{itemize}


To summarize, we analyzed neutrino masses and mixings as the solutions of the
solar neutrino problems and atmospheric neutrino deficits based on a mass
matrix ansatz. Recent Super-Kamiokande results for atmospheric neutrino
showed that the muon neutrino deficits may be explained by 
large mixing between $\nu_{\mu}\leftrightarrow\nu_{\tau}$.
The solar neutrino problems have three possible solutions: small mixing
MSW, large mixing MSW, and just-so vacuum oscillation solutions.
Depending on the solutions to the solar neutrino problems,
we have three possible mixing matrices in the three-neutrino scheme.
For each case we investigated the neutrino mass bounds, 
the magnitudes of mixing matrix elements, and possible nonvanishing 
CP-violating rephasing-invariant quantity $J^l_{\mbox{\tiny CP}}$. 
We conclude that LMA-MSW and VO solutions may come from the
mass matrix ansatz with the similar mass hierarchy:
$m_1 \simeq m_2 \ll m_3$. 
And $J^l_{\mbox{\tiny CP}}$ also has almost the same
magnitude in the two cases, and may reach values up to values of 0.012. 
The origin of SMA-MSW
solution may be attributed to the mass hierarchy: $m_1 \ll m_2 \ll m_3$ with our
mass matrix ansatz. The magnitude of $J^l_{\mbox{\tiny CP}}$
depends on the CP  phase $\delta^l$. In the range of $0\leq \delta^l \leq
\pi/2$, the value of $J^l_{\mbox{\tiny CP}}$ may be up to $1.3\times
10^{-3}$, which is small compared to LMA-MSW or VO solution. 
In $\pi/2 < \delta^l < \pi$, $J^l_{\mbox{\tiny CP}}$ is even more suppressed.

\section*{\bf Acknowledgments}
\noindent
We thank G. Cvetic and S. Pakvasa for careful reading of the manuscript 
and their valuable comments.  The work of C.S.K. was supported 
in part by  Grant No. 1999-2-111-002-5 from the Interdisciplinary 
research program of the KOSEF, in part by the BSRI Program of MOE, 
Project No.  99-015-DI0032, and in part by the KRF Sughak-research 
program, Project No. 1997-011-D00015.
The work of J.D.K. was supported in part by a grant
from the Natural Science Research Institute, 
Yonsei University in the year 1999.

\begin{figure}
\begin{center}
\mbox{\epsfig{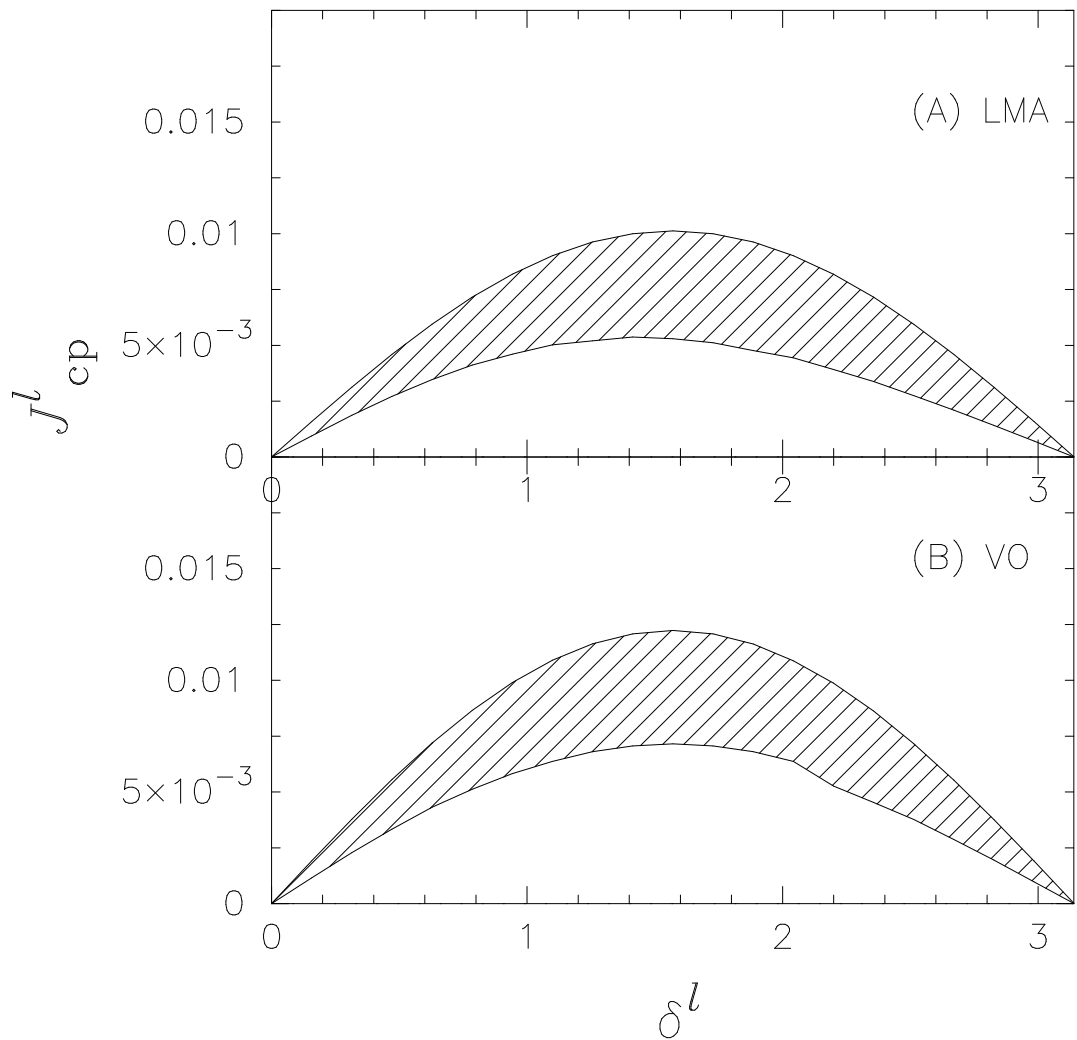}}
\end{center}
\caption{The leptonic CP-violating quantity,
$J^l_{\mbox{\tiny CP}}$, as a function of the CP phase, $\delta^l$.
The hatched regions are allowed by 95\% C.L. mass squared difference 
and mixing angle from the solar and atmospheric neutrino data.
We consider the possible solutions for the solar neutrino problem:
(A) Large Mixing Angle (LMA) MSW,
and (B) Vacuum Oscillation (VO). }
\end{figure}

\pagebreak
\begin{figure}
\begin{center}
\mbox{\epsfig{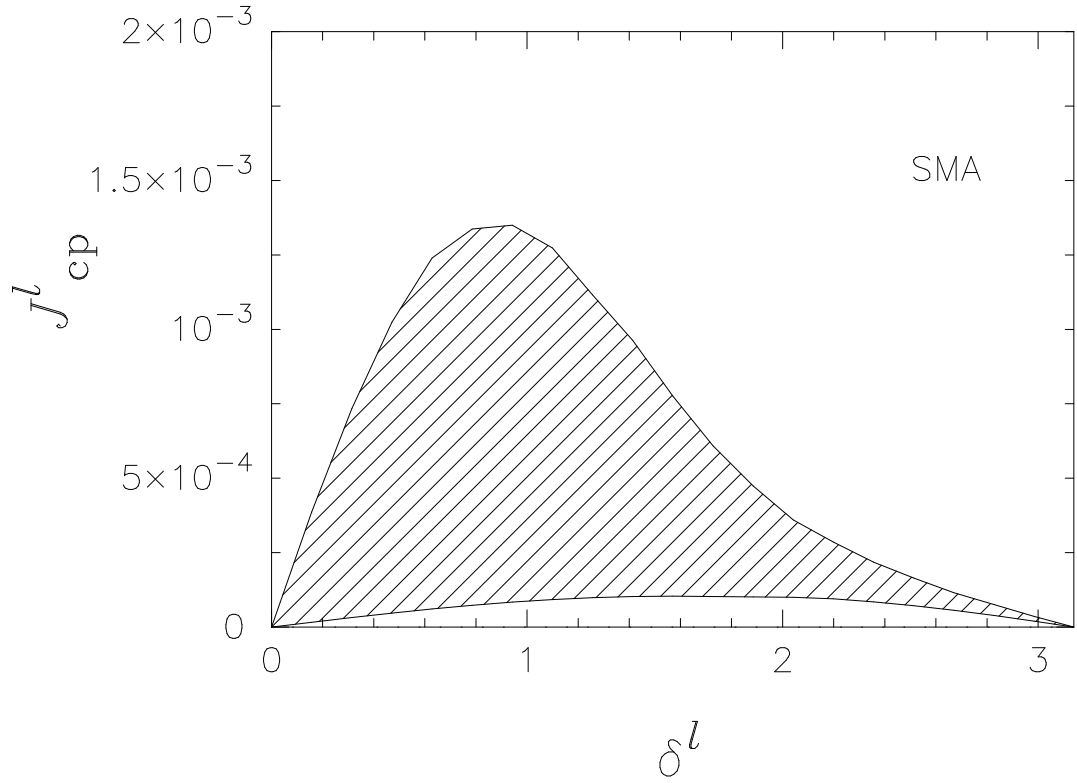}}
\end{center}
\caption{The leptonic CP-violating quantity,
$J^l_{\mbox{\tiny CP}}$, as a function of the CP phase, $\delta^l$.
The hatched region is allowed by 95\% C.L. mass squared difference
and mixing angle from the solar and atmospheric neutrino data.
We consider one of possible solutions for the solar neutrino problem:
Small Mixing Angle (SMA) MSW.
}
\end{figure}

\end{document}